\documentclass{aastex}
\usepackage{lscape}

\newcommand{\ltsima}{$\; \buildrel < \over \sim \;$}
\newcommand{\simlt}{\lower.5ex\hbox{\ltsima}} 
\newcommand{\gtsima}{$\; \buildrel > \over \sim \;$}
\newcommand{\simgt}{\lower.5ex\hbox{\gtsima}} 

\newcommand{\sax}{{\emph{Beppo}SAX} }

\newcommand{\lum}{erg~s$^{-1}$}
\newcommand{\flux}{erg~cm$^{-2}$~s$^{-1}$}
\newcommand{\nh}{cm$^{-2}$}
\newcommand{\norm}{photons~keV$^{-1}$~cm$^{-2}$~s$^{-1}$}

\newcommand{\kev}{\,\mbox{\scriptsize keV}}
\newcommand{\sorg}{Arp~299 }

\begin{document}

\title{An enshrouded AGN in the merging starburst system Arp~299 revealed 
by {\it Beppo}SAX}

\author{R. Della Ceca$^1$, L. Ballo$^1$, F. Tavecchio$^1$, L. Maraschi$^1$, 
P.O. Petrucci$^{1,2}$, 
L. Bassani$^3$, M. Cappi$^3$, M. Dadina$^3$, A. Franceschini$^4$, 
G. Malaguti$^3$, 
G.G.C. Palumbo$^5$ and M. Persic$^6$ 
}

\affil{$^1$ Osservatorio Astronomico di Brera, via Brera 28, 20121 Milan, Italy.
rdc@brera.mi.astro.it}

\affil{$^2$ Laboratoire d'Astrophysique, Observatoire de Grenoble, BP 53X, 38041 Grenoble Cedex, France}

\affil{$^3$ Istituto TeSRE/CNR, via Gobetti 101, 40129, Bologna, Italy}

\affil{$^4$ Dipartimento di Astronomia, Universit\`a degli Studi di Padova, vicolo 
dell'Osservatorio 2, 35122 Padua, Italy}

\affil{$^5$ Dipartimento di Astronomia, Universit\`a degli Studi di Bologna, via Ranzani 1, 40127 Bologna, 
Italy}

\affil{$^6$ Osservatorio Astronomico di Trieste, via G.B. Tiepolo 11, 34131 Trieste, Italy}

\begin{abstract} Using a long ($\simeq$150 ksec), broad-band (0.1--40 keV) \sax
observation of the merging  starburst system Arp~299 (=IC~694 + NGC~3690)
we found the first unambiguous evidence of the 
presence of a deeply buried ($N_H \simeq 2.5\times 10^{24}\:$
\nh) AGN having  an intrinsic luminosity of $L_{0.5-100\:\kev} 
\simeq 1.9 \times 10^{43}\:$\lum. 
The X-ray spectral properties of this AGN are discussed in detail 
as well as the thermal component detected at soft X-ray
energies which, most likely, is associated with the starburst.
\end{abstract}


\keywords{galaxies: active -- galaxies: individual (IC~694, NGC~3690)
-- galaxies: groups (Arp~299)
-- galaxies: nuclei 
-- galaxies: starburst -- X-rays: galaxies}

\section{Introduction}

Studies of active objects at IR and X-ray wavelengths 
indicate that star-formation and AGN activity may be related (Fadda~et~al.~2002).
The triggering mechanism for both phenomena could be the interaction or the
merging of gas-rich galaxies.
This generates fast compression of the
available gas in the inner galactic regions, 
causing both the onset of a major starburst and the fueling of  
a central black hole raising the AGN activity
(see Combes, 2001 for a recent review). 
These two processes may proceed with different time scales 
and once initiated  they probably have different lifetimes.
Therefore a detailed study of the relative importance of starburst and AGN 
activity in the nearest (and brightest) objects
may have profound impact on the understanding of the evolution and the fate
of the gas in the interaction process, on the
modeling the cosmological evolution of both AGNs  and starburst  objects
(e.g. Franceschini, Braito and~Fadda~2002), and 
on the evaluation of the contribution of these sources  
to the energy density of the  Universe 
(e.g. Fabian~et~al.~1998).

The concomitant AGN and starburst activity is 
expected to happen in a high-density  medium ($N_H \geq 10^{23-24}\:$\nh),
characterized by high dust extinction of the UV-optical flux and strong
photoelectric absorption of the soft X-rays (e.g. Fabian et al. 1998). 
Therefore,  
the study of these active phases in galaxies becomes very difficult.
Specific examples are NGC~6240 and NGC~4945. 
Both objects have been classified as LINER and/or starburst galaxies
on the basis of  optical (Veilleux~et~al.~1995)
and mid-/far-IR (Genzel~et~al.~1998) spectroscopy.
For both objects, however, the \sax  PDS observations at $E>10\:$keV have 
clearly revealed the presence of a deeply buried AGN 
($N_H \simeq$ few $\times 10^{24}\:$\nh)
with a QSO-like intrinsic luminosity in the case of 
NGC~6240 ($L_{0.5-100\:\kev} \sim 6\times 10^{44}\:$\lum; Vignati~et~al.~1999) 
and a Seyfert-like luminosity in the case of NGC~4945  
($L_{0.5-100\:\kev} \sim 7\times 10^{42}\:$\lum; Guainazzi~et~al.~2000).
These two examples clearly show that optical and mid-/far-IR
spectroscopy may not be sufficient to disentangle starburst activity from 
AGN activity, which is actually best probed in the hard 
($E > 6\:$keV, 
in order to sample also the Fe K$\alpha$ line) X-ray energy 
band.

To shed light on the starburst-AGN connection and its occurrence we have 
started
a systematic and objective investigation in hard ($E >6\:$keV) X-rays of
{\it a flux-limited sample of IRAS galaxies}.  The sample 
consists
of 28 galaxies selected from the {\it IRAS Cataloged Galaxies and 
Quasars} (see http://irsa.ipac.caltech.edu/) 
as having $f_{60\:\mu\mbox{\scriptsize m}} > 50\:$Jy
or $f_{25\:\mu\mbox{\scriptsize m}} > 10\:$Jy. We 
stress here that no other selection criteria 
(e.g. established presence of an AGN, luminosities, IR colours, etc.) 
have been applied to the sample definition. 

In this letter we present and discuss \sax observations of Arp~299,
a merging system (composed of IC~694 and NGC~3690, but simply 
quoted as NGC~3690 in the IRAS catalogue) located at D$\;=44\:$Mpc 
(z$\;=0.011$; Heckman~et~al.~1999 - H99), spectroscopically classified 
as starbursting from optical (Coziol~et~al.~1998) and mid-/far-IR 
(Laurent et al. 2000) observations.
Its total FIR ($43-123 \mu$m) luminosity, 
$2.86 \times 10^{11}\:L_{\odot}$ (following 
the recipe in Helou et al. 1988 and using the {\it IRAS Faint 
Source Catalogue} fluxes, Moshir et al. 1990), 
dominates the bolometric luminosity.
As shown in this letter the \sax observations
unveil a strongly absorbed ($N_H \simeq 2.5\times 10^{24}\:$\nh) AGN with an
intrinsic (unabsorbed) luminosity of $L_{0.5-100\:\kev} \simeq 1.9\times 10^{43}\:$\lum:  
{\bf this is the first unambiguous evidence of the presence of an AGN in 
the interacting system Arp~299}.
For a detailed description of the system and for a 
summary of previous multiwavelength observations see
Zezas, Georgantopoulos and~Ward (1998; ZGW98),  
H99, 
Hibbard and Yun (1999),  
Alonso-Herrero et al. (2000),
and Charmandaris, Stacey and~Gull~(2002).
In this paper we assume $H_0 = 75\:$~km~s$^{-1}$~Mpc$^{-1}$, and 
$q_0 = 0.5$.

\section{Observations and data reduction}

\sorg was observed by \sax (Boella et al. 1997) in 2001 December 14~-~18
for about 150 ksec.
In this letter we use data collected from 
the Low Energy Concentrator Spectrometer (LECS),
the Medium Energy Concentrator Spectrometer (MECS)
and the Phoswich Detector System (PDS).

The cleaned and linearized data produced by the \sax Scientific Data
Center (SDC, see http://www.asdc.asi.it/bepposax) have been
analyzed using standard software packages (XSELECT v1.4, FTOOLS v4.2 and 
XSPEC 
v11.0). At the best spatial resolution of the \sax instruments 
($\sim 2^{\prime}$), \sorg is not 
resolved;
no significant source flux variation was detected over the whole observing 
period. In the PDS field of view ($\sim 60^{\prime}$ radius)
there are no known and bright (2--10~keV) X-ray sources 
except Arp~299.
In the ROSAT All Sky Survey catalogue we found only two QSOs 
within 60$^{\prime}$ from Arp~299; given their lower flux  
and off-axis angle we can exclude them as sources 
of contamination in the PDS energy range.
 

To maximize statistics and $S/N$, the LECS and MECS source counts were
extracted from a circular region of $4^{\prime}$ radius;
background counts were extracted from high-Galactic
latitude ``blank'' fields (provided by the \sax SDC). The
PDS spectrum extracted with the standard pipeline (with the rise-time 
correction
applied) was provided directly by the \sax SDC;
the simultaneously measured off-source background was used.
In our analysis we took into account only data in the 
0.1--4~keV and 1.8--10~keV energy range for the 
LECS and MECS respectively 
(as suggested in the \sax Cookbook, Fiore~et~al.~1999).
Spectral channels corresponding to energies  10--40~keV have
been used for the PDS data. LECS and MECS (PDS) source counts have been
rebinned to have a $S/N>3$ ($S/N>2$) in each energy bin.
Standard calibration files released by the \sax SDC (September 1997 version)
have been used.
The LECS to MECS and PDS to MECS normalization factors 
were allowed
to vary in the ranges proposed by the \sax Cookbook.
Net exposure times (count rates) are:
63830~s ($0.0087\pm 0.0004\:$cts/s) for the LECS;  
177050~s ($0.0124\pm 0.0003\:$cts/s) for the MECS;  
75490~s ($0.0852\pm 0.0187\:$cts/s) for the PDS.

\section{Spectral analysis}

Single-component models
\footnote{All models discussed here have been filtered 
through a foreground Galactic 
absorption of $N_{H,Gal}=9.92\times10^{19}\:${\nh} 
(Dickey and~Lockman~1990). The thermal component(s) have been 
modelled
with  the MEKAL model. All quoted errors 
are at the 90\% confidence level for 1 parameter of interest ($\Delta \chi
^2=2.71$).}
do not provide an adequate description of the 
broad-band
(0.1--40~keV) spectrum of Arp~299. A single-temperature thermal model or a 
single
unabsorbed power-law model are both rejected at a confidence level greater 
than $99.9\%$.
Figure~\ref{fig:unabspl} shows the ratio between \sax LECS, MECS and 
PDS  data
and the best-fit unabsorbed power-law model ($\Gamma \sim 1.9$). 
A bump at $\sim 0.8\:$keV, a line-like feature 
at $\sim 6.4\:$keV, and a big bump in the PDS energy range
(10--40~keV) are clearly evident.  
The residuals suggest the  presence  of a soft thermal 
component around 0.8~keV, 
which is a characteristic signature in all known starburst galaxy 
X-ray spectra (e.g. Dahlem, Weaver and~Heckman~1998).
{\bf The simultaneous occurrence of a strong 
line-like structure at $\sim \mathbf{6.4}\:$keV and of a big bump in
 the PDS energy range is the distinctive spectral signature of an obscured AGN} 
(Matt et al. 2000).

Given these features, we have performed a broad-band fit which includes:
{\it a)} a soft thermal model;
{\it b)} an unabsorbed power law with photon index $\Gamma$;
{\it c)} a Gaussian emission line at $\sim 6.4\:$keV, and 
{\it d)} an absorbed\footnote{Given that the intrinsic $N_H$ is expected 
to be high ($\sim 10^{24}\:$\nh) we have 
used the PLCABS model (Yaqoob~1997), 
which describes the X-ray transmission correctly taking into account 
Compton scattering.} 
power law having the same photon index $\Gamma$.
The combination of the spectral components 
{\it b)} and {\it d)} is usually called 
``leaky-absorber'' model where the absorbed power law (accounting 
here  for the
PDS energy range)  represents the ``first order'' AGN continuum emerging after 
transmission through an
obscuring cold  medium (torus? starburst related dust?).  
The unabsorbed power law (accounting here
for the
MECS  energy range) represents the primary AGN spectrum scattered into the 
line of sight 
by a warm, highly ionized gas located outside the absorbing medium.
Furthermore, the residuals (not reported here) still show a clear line-like excess 
at $\sim 3.5\:$keV, which has been modelled with a Gaussian emission line.
The results of this best-fit model
\footnote {A simpler model described by a power law plus a thermal 
component and a Gaussian line at $\sim 6.4\:$keV is rejected by the 
data at a confidence level greater than $99\%$ ($\chi^2/$d.o.f=162.1/122).}
($\chi^2/$d.o.f=104.2/118) are
reported in  Table~\ref{tab:saxfit} and are shown in 
Figure~\ref{fig:model}.

\section{Discussion}

\subsection{The AGN component}

The \sax data provide unambiguous evidence of the presence of AGN 
emission from Arp~299, 
with an intrinsic (i.e. unabsorbed) luminosity of
$L_{0.5-100\:\kev} \simeq 1.9\times10^{43}\:$\lum.

The intrinsic power-law photon index ($\Gamma = 1.79^{+0.05}_{-0.03}$) is  
very similar to that usually observed in Seyfert~1 galaxies (Nandra~et~al.~1997). 
The primary continuum is highly absorbed ($N_H = 2.5\times 10^{24}\:$\nh,  with
a 99\% confidence level lower limit  of $\sim  1.76\times 10^{24}\:$\nh)
and can escape from the absorber only at energies above 10 keV:
this indicates that a
deeply buried ``Compton-thick AGN'' lies in the interacting system Arp~299.
The scattered flux fraction (given here as the ratio between the normalizations
of the unabsorbed and absorbed power law components)
implied from the fits is $(5\pm 3)\%$,  
similar to that usually found in Seyfert~2 galaxies 
(e.g. Turner~et~al.~1997).

At the spectral resolution of the \sax MECS, the Gaussian line at $\sim 6.4\:$keV 
is unresolved with a rest frame energy position of $6.42\pm0.13\:$keV and an
observed 
equivalent width (EW) of $636^{+236}_{-270}\:$eV. This value is significantly 
larger than
that observed in Seyfert~1 galaxies  (EW$\:=230\pm 60\:$eV, 
Nandra~et~al.~1997)
but is similar to that measured in Seyfert~2 galaxies 
(e.g. Bassani~et~al.~1999).
The measured position is not consistent with He-like ($6.7\:$keV) or 
H-like ($6.96\:$keV) Fe, as expected 
if the line were produced inside  the
warm, highly ionized gas which scatters the primary AGN spectrum. 
However it is
consistent with the low ionization Fe-K$\alpha$ line from cold material. 
This in turn leads to two possibilities: the line 
could be produced by a reflected and/or a transmitted 
component\footnote{It should be recalled that Fe-K$\alpha$ emission is also typical of
High Mass X-ray 
Binaries (HMXBs) spectra 
(White~et~al.~1983). However, its equivalent width, either observed directly in 
HMXBs (typically EW $\sim 0.3\:$ keV: see White~et~al.~1983) or inferred for galaxies 
with X-ray emission dominated by HMXBs (Persic \& Rephaeli 2002), is inconsistent 
with the large value observed here (if the 
emission has to be completely accounted for by HMXBs).}. 
The first possibility however is unlikely since a significant reflection 
component cannot be easily accommodated  within the \sax data
(using the PEXRAV model, the 90\% upper limit on the reflection 
fraction is 0.07).
The second possibility invokes
the production of the cold Fe-K$\alpha$ line by  transmission through the same 
absorbing
medium that affects the primary AGN  continuum:
indeed, the EW measured with
respect to the transmitted  component, $\sim 7\:$keV, is consistent 
(within the errors on the  intrinsic $N_H$) with what is expected 
from transmission (see e.g.
Leahy and ~Creighton~1993), making this possibility highly plausible.
We note here that the absorbing medium in composite starburst-AGN galaxies, 
such as Arp~299,  is not univocally associated to the putative absorbing  torus
but the nuclear starburst itself  could be a significant 
source of absorption (Fabian~et~al.~1998;  Levenson, Weaver
and~Heckman~2001). 
The measured EW (with respect to the transmitted  component) for Arp~299 
line is a factor  $\sim$ 3-4 above the prediction from the model in 
Ghisellini~et al.~(1994) obtained using  a simple toroidal geometry, which 
is a strong hint towards assuming the above argument apply.

Finally, the unresolved Gaussian line at $3.38^{+0.16}_{-0.13}\:$keV 
(EW$\simeq 125\:$eV) is consistent  with the ArXVIII K$\alpha$ line. 
This line, which originates in the hot scattering medium itself 
(e.g. Netzer, Turner and~George~1998), has been observed also in the 
Compton-thick Seyfert~2 galaxy Circinus (EW$\sim 60$ eV; Sambruna~et~al.~2001).
Based on the flux ratio between the ArXVIII K$\alpha$ line and the Fe He-like line
(which should be produced inside the same hot scattering medium) observed in Circinus, 
in Arp~299
we would expect an Fe He-like line with EW $\sim\:$400~eV, 
a value consistent (within a factor of 2) 
with the derived~90\% upper limit.

\subsection{The Starburst Component}

The best-fit soft X-ray thermal component gives a $kT\simeq
0.86\:$keV. This value is consistent, within the errors, with that previously 
obtained by 
ZGW98 and H99 using ROSAT PSPC plus ASCA data.
It is also similar to that usually found in other well studied starburst
galaxies (e.g. Dahlem, Weaver and~Heckman~1998).
This thermal emission likely originates from the interaction between
hot, low-density galactic winds and cold, high-density ISM  
(see H99). 

Starburst galaxies are also characterized by a hard 2--10~keV  spectral 
component interpreted as being either thermal with $kT\sim 5-10$ keV 
or non-thermal with $\Gamma \sim 1.5-2$ (e.g. Dahlem, Weaver and~Heckman~1998).
Such hard component is likely to be the integrated emission of
X-ray binaries, and can be modelled by means of a 
cutoff power-law, $f(E) \propto E^{-\Gamma} e^{-E/kT}$ 
(CPL model, see Persic and~Rephaeli~2002).

In order to check to what extent the 2--10~keV emission   
from \sorg can be related to the starburst, in our model 
we have replaced the scattered AGN component (labelled as {\it b} in section 3) 
with:
{\it a)} a thermal component or 
{\it b)} the absorbed CPL model under the hypothesis
that the main contribution is due to X-ray binaries.
Both alternatives are statistically viable,  with:
{\it a)} the thermal model with $kT =6.58^{+1.74}_{-1.18}$ or 
{\it b)} the unabsorbed CPL model with $\Gamma = 1.65 \pm 0.15$ ($kT$
fixed at 8~keV).
In both cases, the spectral parameters of the soft X-ray component
and of the absorbed  AGN component remain consistent (within the errors) with those 
reported in Table~\ref{tab:saxfit}.
Since the inferred 2--10~keV luminosity of both the thermal and the CPL components 
is $\sim 2 \times 10^{41}\:$\lum  
(reasonable for a bright FIR galaxy like \sorg) 
we cannot rule out that at least part of the observed 2--10~keV  X-ray luminosity is due 
to the X-ray binaries directly associated with the starburst 
(which may also contribute to the observed Fe-K$\alpha$ emission)
\footnote {If this is the case, than the AGN scattered flux fraction reported in 
section 4.1, $(5\pm 3)\%$, should be considered as an upper limit.} 
.
This conclusion was reached by 
ZGW98 and H99 using ASCA data.
However the ASCA statistics and energy coverage
were insufficient to clearly detect the 6.4~keV Fe-K$\alpha$ emission line and 
to reveal the high  energy bump
\footnote{We have directly verified that the archival 
{\it ASCA} data are consistent with the best fit spectrum reported 
in Table~\ref{tab:saxfit} 
and are not deep enough to clearly detect the 
6.4~keV Fe-K$\alpha$ emission line.}. 
The new broad-band \sax data instead clearly require the presence of an absorbed  
AGN component.

\section{Conclusions}

In a \sax observation of the merging starburst system Arp~299 
the 0.1--40~keV data coverage clearly 
reveals, for the first time in this system, 
the presence of a deeply buried ($N_H \simeq
2.5\times 10^{24}\:$\nh) AGN with an intrinsic 
(i.e. unabsorbed)  
luminosity of $L_{0.5-100\:\kev}
\simeq 1.9\times 10^{43}\:$\lum.
This AGN component was missed in previous 
multiwavelength observations.
Assuming a Galactic standard value of $E_{B-V}/N_H = 1.7\times 10^{-22}$ mag 
cm$^2$ (Bohlin~et~al.~1978) and the measured $N_H$ value we argue that the 
AGN is completely absorbed in the optical and  IR band  ($A_V > 1000$ 
mag!!).

The intrinsic AGN's X-ray luminosity is a factor $\sim$ 50 less than 
$L_{FIR}$ ($\sim 10^{45}\:$\lum).
Thus the total FIR luminosity of the system cannot be 
entirely associated to the AGN  
even assuming an AGN UV~luminosity 
a factor  $\sim\:$10 greater than the X-ray luminosity (as
observed in QSOs, e.g. Elvis~et~al.~1994).
This suggests that the bulk FIR emission  
may be due to the starburst, in agreement with the results
obtained by Laurent et al. (2000) using 5-16$\mu$m ISOCAM data.
This conclusion may also be supported by 
the fact that the ratio of the soft X-ray 
thermal emission (likely associated with the starburst) 
to the FIR emission ($\sim 10^{-4}$ in the case of 
Arp~299) is similar to other 
``bona fide'' starburst galaxies such as M82 ($\sim 6 \times 10^{-5}$)
and  NGC~253 ($7\times 10^{-5}$).

 
A question unsolved at the present stage is the location of the AGN
inside the interacting system Arp~299.
At the spatial resolution of the \sax instruments  
we are collecting photons from
Arp~299 (= IC694 + NGC 3690) as well as from the nearby galaxies Arp296 
and MCG+10-17-2a.  It is safe to exclude the last two as the origin of 
the hard X-ray emission since they are very weak sources in both the 
radio and in the IR, and they were not detected in soft X-rays by ZGW98.
On the other hand both IC694 and NGC~3690 could be the host of the 
AGN.
We have recently retrieved the public {\it Chandra} and XMM-Newton data for this 
system. From a preliminary analysis a strong line at $\sim 6.4$ keV is clearly 
present in NGC~3690 and maybe also in IC694, suggesting the possible presence 
of two AGNs. The results of this analysis, which will also include an 
estimate of the binaries contribution to the 2-10 keV emission and a detailed 
comparison of the X-ray emission with that at other wavelenghts, 
will be reported in a forthcoming paper.

\acknowledgements


We thank V. Braito, P. Severgnini, T. Maccacaro, G. Trinchieri 
and the anonymous referees for useful comments.
Partial financial support from ASI (I/R/073/01) 
and MIUR (Cofin-00-02-004) is acknowledged.

\clearpage

\centerline{ \bf Figure Captions}

\figcaption[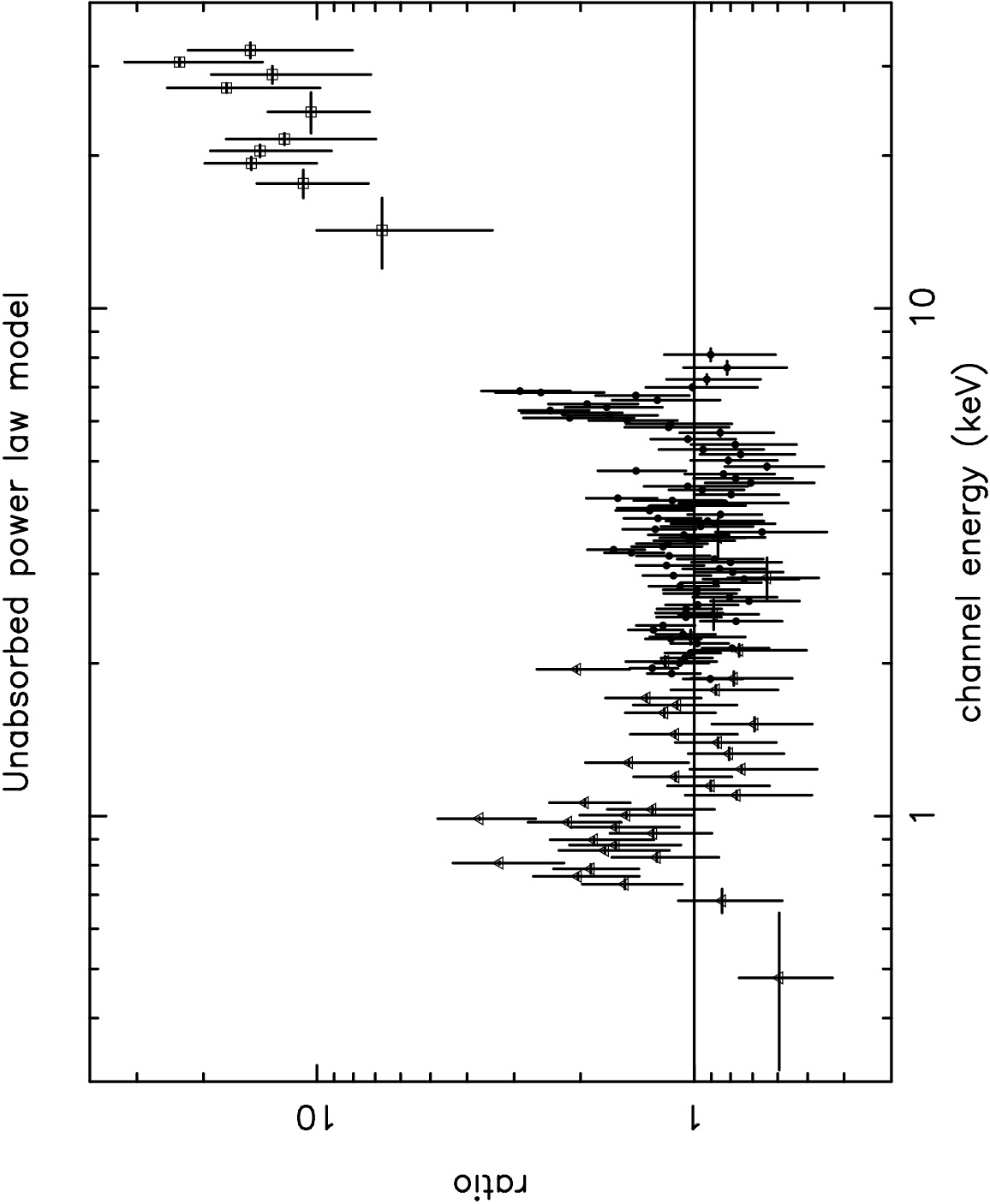]{Ratio between the unabsorbed power-law model and the 
\sax LECS (open triangles),
MECS (filled circles) and PDS (open squares) data. The PDS to MECS 
normalization factor is constrained to vary within the range indicated by the 
\sax Cookbook (Fiore~et~al.~1999).
\label{fig:unabspl}}

\figcaption[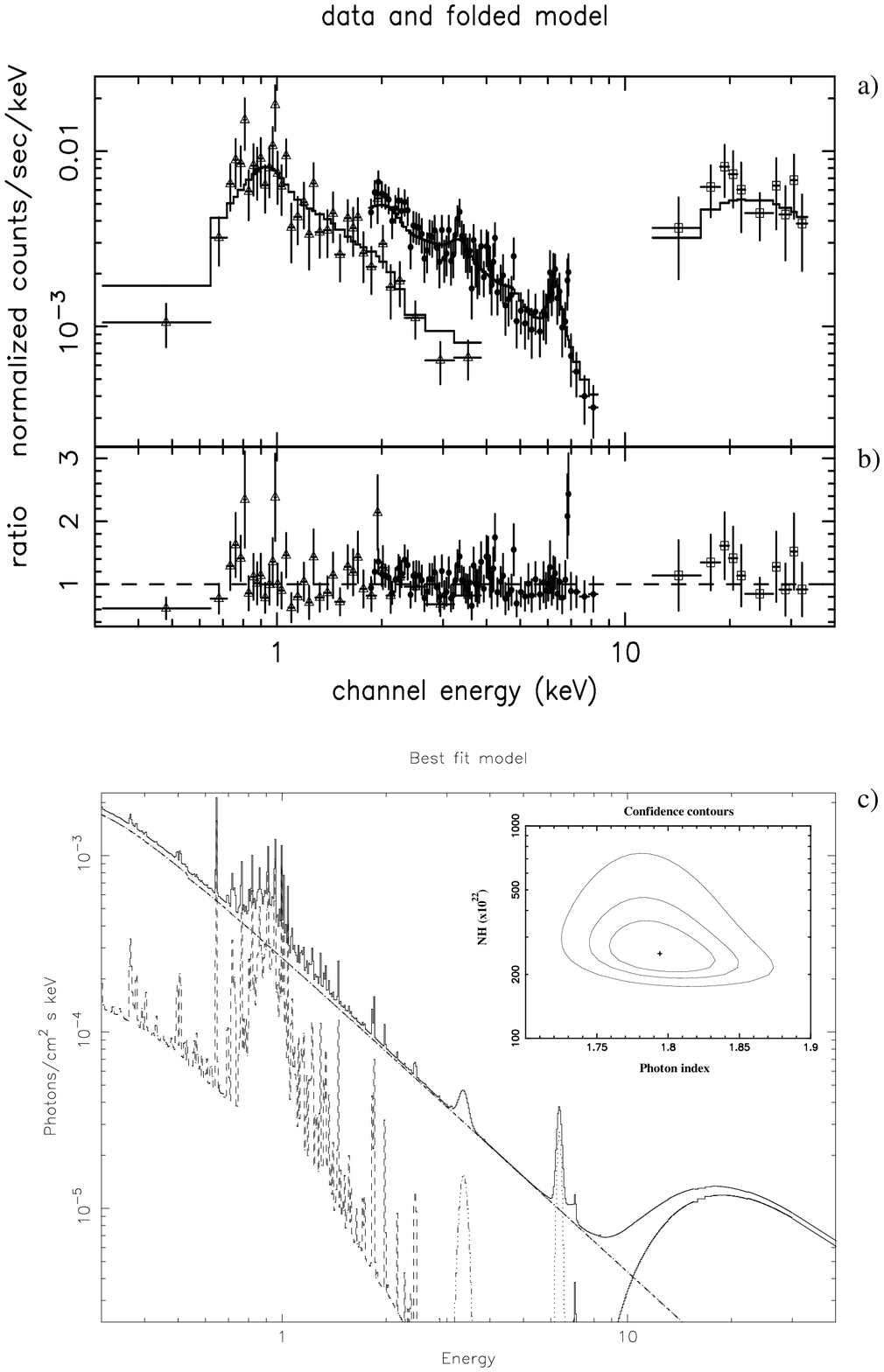]{Data fit with a MEKAL thermal component, the
``leaky-absorber'' continuum and two narrow Gaussian lines 
(see Table~\ref{tab:saxfit}). {\it
Panel a}: LECS (open triangles), MECS (filled circles) and PDS (open squares) 
folded spectra and data; {\it panel b}: ratio 
between the
data and the best-fit model; {\it panel c}: the unfolded model. {\it Panel c insert:}
confidence contours ($68\%$, $90\%$ and $99\%$ confidence level for 
two
interesting parameters) for the power law photon index and the absorbing 
column density.
\label{fig:model}}

\clearpage

\begin{landscape}
\begin{table*}
 \begin{center}
 \caption{Results of the spectral fit (LECS+MECS+PDS): MEKAL thermal component~+~absorbed and scattered power law~+~narrow Gaussian lines ($\chi^2/$d.o.f=104.2/118), and unabsorbed X-ray luminosities}\label{tab:saxfit}
 \vspace{1cm}
 \hskip -2truecm
  \begin{tabular}{rc|ccc|cccc}
    \hline
    \hline
   & & $kT$ (keV)/$\Gamma$/$E_{\mbox{\scriptsize line}}$ (keV) & Norm & $N_H$/EW (eV) & \multicolumn{3}{c}{Luminosity (\lum)} & \\
   & & & & & 0.5--2 keV & 2--10 keV & 10--100 keV & \\
   \hline
   \multicolumn{1}{r|}{SB} & MEKAL$^{(a)}$ & $0.86^{+0.24}_{-0.19}$ & $10.29^{+5.98}_{-4.63}$$^{(b)}$ & & 1.12$\times 10^{41}$ & 7.41$\times 10^{39}$ & 5.36$\times 10^{35}$ & \\
    \hline
   \multicolumn{1}{r|}{} & Absorbed P. L. & $1.79^{+0.05}_{-0.03}$$^{(c)}$ & $590^{+523}_{-169}$$^{(d)}$ & $2.52_{-0.56}^{+1.39}$$^{(e)}$ & & & & \\
   \multicolumn{1}{r|}{AGN} & Unabsorbed P. L. & $1.79^{+0.05}_{-0.03}$$^{(c)}$ & $26.93_{-3.93}^{+4.29}$$^{(d)}$ & & 3.09$\times 10^{42}$ & 4.93$\times 10^{42}$ & 1.06$\times 10^{43}$ & \\
   \multicolumn{1}{r|}{} & Lines (rest frame) $^{(f)}$ & $3.38^{+0.16}_{-0.13}$ & $0.39_{-0.26}^{+0.30}$$^{(g)}$ & $125^{+98}_{-84}$ & & & & \\
   \multicolumn{1}{r|}{} & & $6.42^{+0.13}_{-0.13}$ & $0.68_{-0.26}^{+0.29}$$^{(g)}$ & $636^{+236}_{-270}$ & & & & \\
   \hline
    \multicolumn{9}{c}{}\\
     \multicolumn{9}{l}{\footnotesize NOTE: Errors are quoted at the 90\% confidence level for 1 parameter of interest ($\Delta \chi ^2=2.71$). The LECS to MECS and PDS to MECS}\\
     \multicolumn{9}{l}{\footnotesize normalization factors ($0.7$ and $0.95$, respectively) are consistent with the known differences in the absolute calibration of the instruments.}\\
     \multicolumn{9}{l}{\footnotesize The total observed fluxes of Arp~299 are $8.16\times10^{-13}\,$\flux (0.5--2 keV), $1.13\times10^{-12}\,$\flux (2--10 keV) and}\\
     \multicolumn{9}{l}{\footnotesize $3.23\times10^{-11}\,$\flux (10--100 keV). $^{(a)}$ The metallicity of the thermal component was fixed to the Solar one. $^{(b)}$ In units of}\\
     \multicolumn{9}{l}{\footnotesize $[10^{-19}{/}(4\pi D^2) \int n_e n_H\,dV]\,$@$1\,$keV, where $D$ is the distance of the source in cm, $n_e$ and $n_H$ are the electron and $H$ density in units of} \\
     \multicolumn{9}{l}{\footnotesize cm$^{-3}$, and $V$ is the volume filled by the X-ray emmitting gas in cm$^3$. $^{(c)}$ The two power law photon index have been fitted together.} \\
     \multicolumn{9}{l}{\footnotesize $^{(d)}$ In units of $10^{-5}\,$\norm\,@$1\,$keV. $^{(e)}$ Column density of neutral hydrogen (in units of 10$^{24}$ cm$^{-2}$) in addition } \\
     \multicolumn{9}{l}{\footnotesize to $N_{H,Gal}=9.92\times10^{19}\:$\nh. $^{(f)}$ The lines are unresolved at the spectral resolution of the \sax MECS. }\\
    \multicolumn{8}{l}{\footnotesize  $^{(g)}$ In units of $10^{-5}\,$photons~keV$^{-1}$~cm$^{-2}$~s$^{-1}$ at the energy of the line.}\\
  \end{tabular}
 \end{center}
\end{table*}
\end{landscape}

\newpage

\begin{figure}
 \centerline{\plotone{f1.eps}}
\end{figure}

\newpage

\begin{figure}
 \centerline{\plotone{f2.eps}}
\end{figure}

\end{document}